\documentclass[aps,prb,preprint,showpacs,groupedaddress]{revtex4}
\usepackage{mathrsfs}
\usepackage{graphicx}
\begin{document}

\title{Low frequency elastic wave propagation in 2D locally resonant phononic crystal with asymmetric resonator}

\author{Yongwei Gu}
\author{Xudong Luo}
\email{luoxd@sjtu.edu.cn}
\author{Hongru Ma}
\affiliation{Department of Physics, Shanghai Jiao Tong University,
Shanghai 200240, People's Republic of China}

\date{\today}

\begin{abstract}
The resonance modes and the related effects to the transmission of
elastic waves in a two dimensional phononic crystal formed by
periodic arrangements of a two blocks unit cell in one direction are
studied. The unit cell consists of two asymmetric elliptic cylinders
coated with silicon rubber and embedded in a rigid matrix. The modes
are obtained by the semi-analytic method in the least square
collocation scheme and confirmed by the finite element method
simulations. Two resonance modes, corresponding to the vibration of
the cylinder along the long and short axes, give rise to resonance
reflections of elastic waves. One mode in between the two modes,
related to the opposite  vibration of the two cylinders in the unit
cell in the direction along the layer, results  in the total
transmission of elastic waves due to  zero effective mass density at
the frequency. The resonance frequency of this new mode changes
continuously with the orientation angle of the elliptic resonator.

\end{abstract}


\maketitle

\section{Introduction}
The propagation of elastic or acoustic waves in periodic
heterogeneous materials has received  renewed attention in the last
several years\cite{liu2000science, hirsekorn2004apl,
hirsekorn2006jap, wu2007jpdap, acoustic_lense, goffaux2003apl,
sanchez2002apl, goffaux2003prb, wang2004prb, wang2004prl}. A new
class of composite materials with periodic structures in the elastic
properties has been proposed to realize the sound attenuation at
selective frequency bands in the audible range, from  several
hundred hertz to few kilohertz \cite{liu2000science}. For suitable
periodicity and elastic contrasts between the components,
low-frequency gaps may appear in these so-called phononic crystals
(PCs) in the form of a thin layer, which can lead to promising
applications in sound insulation \cite{hirsekorn2004apl,
hirsekorn2006jap}. The periodic structures can also be applied to
the design of acoustic wave guides or filters \cite{wu2007jpdap,
acoustic_lense}.

It has been demonstrated the improvements on the sound insulation of
the PCs in comparison with the homogenous materials whose ability of
sound insulation is governed by the mass law \cite{goffaux2003apl,
sanchez2002apl}. Specific efforts were made to the identify and
application of the PCs' characteristics  in the low-frequency range.
The so-called first resonance frequency, the local vibration of the
constitution elements of PCs, which is usually a coated sphere or
other simple shapes,
 was thus extensively examined
\cite{goffaux2003prb, wang2004prb, wang2004prl}.

Theoretical methods have been developed in order to understand the
properties of elastic response of PCs and predict new phenomena. A
conceptually simple method named plane wave expansion method (PWE),
which treats the wave equation in the Fourier space
\cite{pwe_method}, has obtained a vast of important results of the
PCs. However, when used in problems with large elastic constant
contrasts, it suffers from slow convergence and heavy computation
efforts.
 On the contrary, the multiple scattering method
\cite{mst}, looks a little bit complicated,  overcomes these
difficulties. The problem with it is  that the method, though very
efficient in simple elementary geometries such as spheres in three
dimensional and cylinders with circular cross section in two
dimensional problems, is not efficient to deal with nonelementary
resonator's geometries. In this case, it is practical to resort to
some computational techniques, such as Multiple Multipole method
(MMP) \cite{imhof1996jasa, imhof2004jasa, geophysics},
 finite difference algorithms \cite{fdtd_method}, lumped
mass method (LM) \cite{wang2004prb}, and  variational method (VM)
\cite{goffaux2003prb}, etc.

In addition to the above mentioned successful methods, mechanical
models have been developed for cylindrical elastic resonators (2D
PCs). The models are based on a point mass connected as a pendulum
or by springs to a rigid or elastic matrix
\cite{goffaux2003prb,hirsekorn2004apl} and are able to predict the
dependence of the resonator on the material constants. These models
are useful to gain insight into the underlying physics. Moreover,
the line profile of the transmittance spectra for the first
resonance peak was analyzed by C. Goffaux \cite{goffaux2002prl}
along this line. Their mechanical model, taking into account the
interaction between propagating waves in the matrix material and
local resonance effects, evidenced the equivalence of resonance
scattering in PCs to Fano's interference phenomena. Quite recently,
complete modal analysis has been performed for single cell with
Finite Element Method (FEM). The resonance frequencies of modal
analysis revealed by FEM are compared with the simulated wave
attenuation peaks. A small phase shift was observed and interpreted
as Fano type interferences \cite{hirsekorn2006jap}. However, since
the methods are based on PC whose repeating unit cell contains only
one resonator connected to a matrix, thus can not treat properties
that associated with local couplings between two nearest resonators,
symmetric or asymmetric.

In this paper, we use a different approach to study a 2D ternary PC
with asymmetric resonators. We investigate resonance modes resulting
from PC with two connected resonators as the unit cell, using
analytic formulation combined with Least Square Collocation Method
(LSCM) \cite{imhof1996jasa,topublish}, whose more advanced version
is known as multiple multipole method. Then the model is simulated
by FEM to confirm the existence of these resonance modes. With
asymmetric arrangement of the two resonators, we identified  a new
resonance mode attributed to the coupling effects of the two
resonators in a unit cell. Wave transmission spectrum is also
calculated by FEM. The line profile exhibited in transmission
spectrum corresponding to this new resonance mode is also discussed.

\section{The Semi-Analytical treatments and Results}\label{part2}

The resonance modes of a spherically symmetric elastic resonator and
its cylindrically symmetric counterpart in two dimension (2D) have
been derived analytically under the assumption of rigid core and
matrix \cite{liu2005prb} in 2005. In 2006, Hirsekorn $et$ $al$
\cite{hirsekorn2006jap}  extended the treatment by removing  the
rigid assumption  and employing the physical material properties.
And the analytic model  were confirmed by the simulation results.
Here, we derive an analytical formulation  describing an asymmetric
2D model, combined with LSCM for boundary conditions.

\begin{figure}
\includegraphics[scale=1]{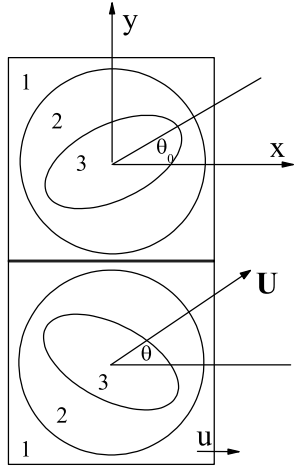}
\caption{The unit cell of the PC in this study, constructed by two
blocks of the ternary material. The two blocks are placed with
mirror symmetry about the middle plane. The side length of each
block is $d=16mm$, the outer radius of the coating is $c=7.5mm$,
short and long axis of the elliptical cylinder are $a=3mm$ and
$b=5mm$, respectively. $\theta$ is the vibration angle, gives the
polarization of the vibration with respect to the $x$ axis,  and
$\theta_0$ is the orientation angle, specifies the angle between the
long axis of the upper elliptic cylinder and the $x$ axis.}
\label{fig:unitpair}
\end{figure}

The model system we studied is the PC with the unit cell as an asymmetric arrangement of two elliptic hard cylinders
 coated with soft silicone rubber and
embedded in a rigid matrix, the cross section of the unit cell is
shown schematically in Fig. \ref{fig:unitpair}. The region $1$ is
the rigid epoxy matrix, region $2$ are the two soft silicone rubber
coatings with their outer shape as circles, and region $3$ are two
hard cylinders of elliptic shape. The geometric and physical
parameters are given in the figure caption.

Suppose a long wave-length elastic wave is traveling along the $x-y$
plane, or, $r-\phi$ plane, perpendicular to the axis of the
cylinder. The cylinder is vibrating around the cylinder axis.
Consider the lower cell of the pair shown in Fig.
\ref{fig:unitpair}, and suppose that the vibration angle is $\theta$
with respect to the positive $x$ direction (the arrow direction in
Fig. \ref{fig:unitpair}), the cylinder will translate as a rigid
body in both $x$ and $y$ direction,
\begin{equation}
-m_3\omega^2U\cos\theta=\int\int_s[n_r\cos\phi\tau_{rr}+(n_\phi\cos\phi-n_r\sin\phi)\tau_{r\phi}-n_\phi\sin\phi\tau_{\phi\phi}]
rdl d\phi \label{motion_x}
\end{equation} and
\begin{equation}
-m_3\omega^2U\sin\theta=\int\int_s[n_r\sin\phi\tau_{rr}+(n_\phi\sin\phi+n_r\cos\phi)\tau_{r\phi}+n_\phi\cos\phi\tau_{\phi\phi}]
rdl d\phi,\label{motion_y}
\end{equation}
respectively. And its rotation described by
\begin{equation}
-I_3\omega^2\phi_o=\int\int_s r
(n_r\tau_{r\phi}+n_\phi\tau_{\phi\phi})rdl d\phi,\label{rotation}
\end{equation}
where $U$ is the displacement of the cylinder, $\phi_o$ is the
amplitude of the rotation, $m_3$ is the mass of the cylinder,
expressed as $\rho_3\pi a b l$, $I_3$  is the inertial of rotation
with respect to the cylinder axis, given by $\rho_3 l \pi a
b(a^2+b^2)/4$, with $\rho_3$ being the mass density of the cylinder,
$a$ and  $b$ are the short and long semi-axis of the elliptical
cross section of the cylinder, and $l$ is its length. Here,
$\tau_{rr}$ and $\tau_{r\phi}$ are  components of stress tensor in
medium $2$ in a cylindrical coordinate system, and $n_r$ and
$n_\phi$ are the $r$ and $\phi$ components of the outer normal unit
vector on the cylinder surface, which are functions of $(r, \phi)$.
  The surface integration on the right hand side of
(\ref{motion_x}) and (\ref{motion_y})  are the forces exerted on the cylinder from  stresses in  the   coating in $x$ and $y$ direction,
and  the right hand side of
(\ref{rotation}) is the toque on the cylinder.

The displacement field in  medium $2$ is described by the elastic wave equation \cite{graff1991}
\begin{equation}
(\lambda_2+2\mu_2)\nabla(\nabla\cdot\mathbf{u})-\mu_2\nabla\times\nabla\times\mathbf{u}+\rho_2\omega^2\mathbf{u}=0,\label{book}
\end{equation}
where $\lambda_2$ and $\mu_2$ are the Lam\'{e} coefficients , and
$\rho_2$ is the mass density of medium $2$. Mirror symmetry about the
$x-y$ plane indicates that the displacement in the lower block is just the
mirror reflection of the upper block. This means that the $y$ direction displacement of the
reflection plane is zero. Thus we only need to obtain the displacement fields in one block, which we choose the upper
block to solve.
The equation of the displacement field in medium $2$ may be
expressed in terms of potential functions as \cite{graff1991}
\begin{equation}
\mathbf{u}=\nabla\Phi+\nabla\times(\Psi\vec{e_z}),
\label{displacement}
\end{equation}
where $\Phi$ and $\Psi$ are scalar potential functions, satisfying
the following scalar wave equations:
\begin{eqnarray}
\nabla^2\Phi+\alpha^2\Phi=0,\label{wave1}\\
\nabla^2\Psi+\beta^2\Psi=0,\label{wave2}
\end{eqnarray}
here $\alpha=\omega\sqrt{\rho_2/(\lambda_2+2\mu_2)}$ and
$\beta=\omega\sqrt{\rho_2/\mu_2}$, denoting the wave numbers of the shear($S$)-wave
and compressional($P$)-wave, respectively. The solutions for $\Phi$ and $\Psi$ may
be written as
\begin{eqnarray}
\Phi=\sum_{k=0}^\infty[A_kJ_k(\alpha r)+B_kN_k(\alpha
r)](C_k\cos k\phi+D_k\sin k\phi), \label{phi}\\
\Psi=\sum_{k=0}^\infty[E_kJ_k(\beta r)+F_kN_k(\beta r)](G_k\sin
k\phi+H_k\cos k\phi),\label{psi}
\end{eqnarray}
where $J_k(x)$ is the Bessel function and $N_k(x)$ the Neumann
function. The angle part in (\ref{phi}) and (\ref{psi}) can be expressed in
linear combinations of $\sin\phi$ and $\cos\phi$, while the
single-valued requirement  indicates that $k$ is an integer.
For practical reasons, the expansion of $\Phi$ and $\Psi$ have to be
truncated after $K$ terms, where index $k\in\{0,1,\cdots, K\}$. The
displacement field is continuous across the boundaries between
medium $1$ and $2$ ($\partial \Gamma_{12}$)
\begin{eqnarray}
u_r|_{r=c}=u\cos\phi,\label{continuity 121}\\
u_\phi|_{r=c}=-u\sin\phi,\label{continuity 122}
\end{eqnarray}
and between medium $2$ and $3$ ($\partial \Gamma_{23}$)
\begin{eqnarray}
u_r|_{\partial \Gamma_{23}}=U\cos\theta\cos\phi +U\sin\theta\sin\phi -r (1-\cos\phi_o),\label{continuity 231}\\
u_\phi|_{\partial \Gamma_{23}}=-U\cos\theta\sin\phi +U\sin\theta\cos\phi +r \sin\phi_o,\label{continuity
232}
\end{eqnarray}
where $u$ is the displacement of the epoxy.

In a local cylindrical coordinate system $(r, \phi, z)$, the strains
due to a displacement $\mathbf{u}$ are expressed as\cite{graff1991}
\begin{eqnarray}
&&\epsilon_{rr}=\frac{\partial u_r}{\partial r},\label{strain displacement relation 1}\\
&&\epsilon_{\phi\phi}=\frac{\partial
u_\phi}{\partial\phi}+\frac{\partial u_r}{\partial r},\label{strain
displacement relation 2}\\
&&\epsilon_{r\phi}=\epsilon_{\phi
r}=\frac{1}{2}\left(\frac{1}{r}\frac{\partial
u_r}{\partial\phi}+\frac{\partial u_\phi}{\partial
r}-\frac{u_\phi}{r}\right). \label{strain displacement relation 3}
\end{eqnarray}
All other components are zero since they involve the $u_z$ component
or cross derivatives with respect to $z$. The stresses are linearly
related to the strains by \cite{graff1991}
\begin{equation}
\tau_{pq}=\lambda\delta_{pq}\sum_i\epsilon_{ii}+2\mu\epsilon_{pq},
\quad\quad p,q\in\{r,\phi\}. \label{stress stain relation}
\end{equation}

The displacement field is determined when the coefficients
$A_k$-$H_k$ are known, which then can be obtained from the boundary
conditions, (\ref{continuity 231}) and (\ref{continuity 232}). Since
the coordinate system used is not coincident with the boundaries of
the core cylinders, the wave equations (\ref{wave1}) and
(\ref{wave2}) are not separable on the boundaries. In principle, the
boundary condition should be satisfied on every point of the
boundaries, in practice, the expansions of (\ref{phi}) and
(\ref{psi}) are truncated at $K$th term so that only a finite number
of independent equations are needed to solve for the unknowns. The
simplest way to solve the problem is to choose as many
representative matching points on the boundaries as the number of
unknown coefficients and solve for the coefficients from boundary
conditions on the
  matching points. However, the solution obtained this way is not necessary to satisfy
the boundary conditions at other points on the boundaries, and worst
of all, the equations resulted in this way may not be solvable. For
example, the resulted equations may be singular, or even may not
independent. This problem is solved by the so called LSCM. In this
method we choose the number of matching points much larger than the
number of unknowns and solve the over determined equations in the
least square sense. The boundary conditions on the matching points
are not exactly satisfied, but the overall deviations from the exact
boundary conditions will be minimal and the solution is "smoother"
in between matching points\cite{imhof1995jasa}. The displacement $U$
and $u$ are predefined symbol constants, thus in our specific
problem we have $8(K+1)$ independent unknowns if the truncation of
the expansion is $K$. On the boundary between region $1$ and $2$, we
choose $L$ matching points and on the boundary between $2$ and $3$,
$M$ matching points are chosen, hence $2(L+M)$ equations. The
inequation $2(L+M)
> 8(K+1)$ should be satisfied.

With the obtained coefficients, the $\tau_{rr}$ and $\tau_{r\phi}$
can be obtained, and then the right-hand side of equations of motion
(\ref{motion_x})-(\ref{rotation}). Finally, the  $U\cos\theta$,
$U\sin\theta$, and $\phi_o$ can be computed from the equations of
motion, in terms of $u$.

We calculated the displacements and other physical properties of the
model for $\theta_0=45^\circ$, the truncation $K=30$, and the
matching points $L=180$, $M=180$, and other  parameters are given in
the caption of Fig. \ref{fig:unitpair}.  Now we discuss our results
in detail.  A resonance rotational mode is found at $f=258$Hz,
denoted as mode $1$. In this mode, the inclusion cylinders
rotationally vibrate within the coating, while the epoxy matrix
remains nearly stationary. This resonance mode could be described by
a model with the shear deformation of coating \cite{wang2004prb}.
However, according to the rule proposed by Wang $et$ $al$
\cite{wang2004prl}, it makes no contribution to wave insulation. The
reason for this is that   the forces to the hosting structure
contributed by the oscillator is zero in this mode.
\begin{figure}[htp]
\centering
\includegraphics[width=0.5\textwidth]{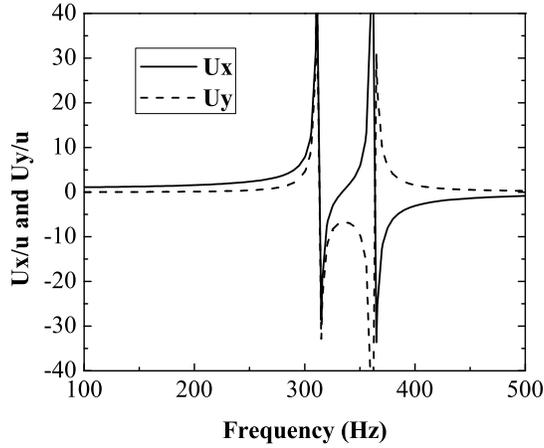} \hfill
\caption{The frequency dependence of displacement for rubber-coated
elliptical cylinders in both $x$ and $y$ directions, denoted by
solid line for $U_x=U\cos\theta$, and dashed line for
$U_y=U\sin\theta$, respectively.}\label{fig:mode124}
\end{figure}
Fig. \ref{fig:mode124} shows the displacement of the core cylinders.
It is observed that there are two resonances  around $f=313$Hz and
$f=363$Hz, denoted as mode $2$ and mode $4$, at these frequencies
the amplitude of the displacements of the cylinders in both
directions are very large and experience a very sharp ($\sim \pi$)
change of phase. By solving for the vibration angle $\theta$, it is
found that the lower  frequency resonance vibration is along the
long semi-axis of the elliptical cylinders and the higher frequency
resonance vibration is along the short semi-axis of the elliptical
cylinders.

The appearance of two different resonance frequencies is not only a
result of the variation of the coating thickness, it also relates to
the contact area between medium 2 and 3 in vibration direction.
Along the short semi-axis the effective thickness is larger than the
one along the long semi-axis, however, the contact area is much
bigger for short semi-axis vibration. These give rise to a higher
resonance frequency in mode 4 than mode 2.
 A detailed description about the two modes without coupling effects can be found in
[\onlinecite{hirsekorn2006jap}]. Between the two resonance
frequencies, resulted from the splitting of the degenerate modes in
symmetric resonator case, the phases of the vibration in $x$ and $y$
direction are reversed. This phase reversion  brings about an
interesting mode, denoted as mode $3$, which demonstrates some
notable physical properties.

\begin{figure}
\centering
\includegraphics[width=0.45\textwidth]{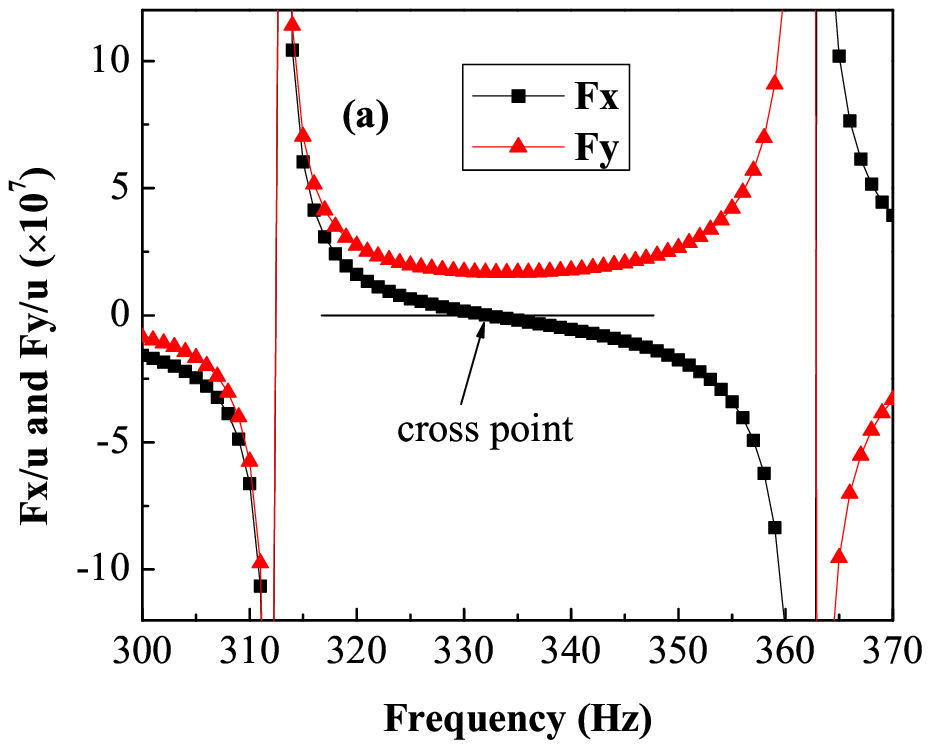}\hfill
\includegraphics[width=0.45\textwidth]{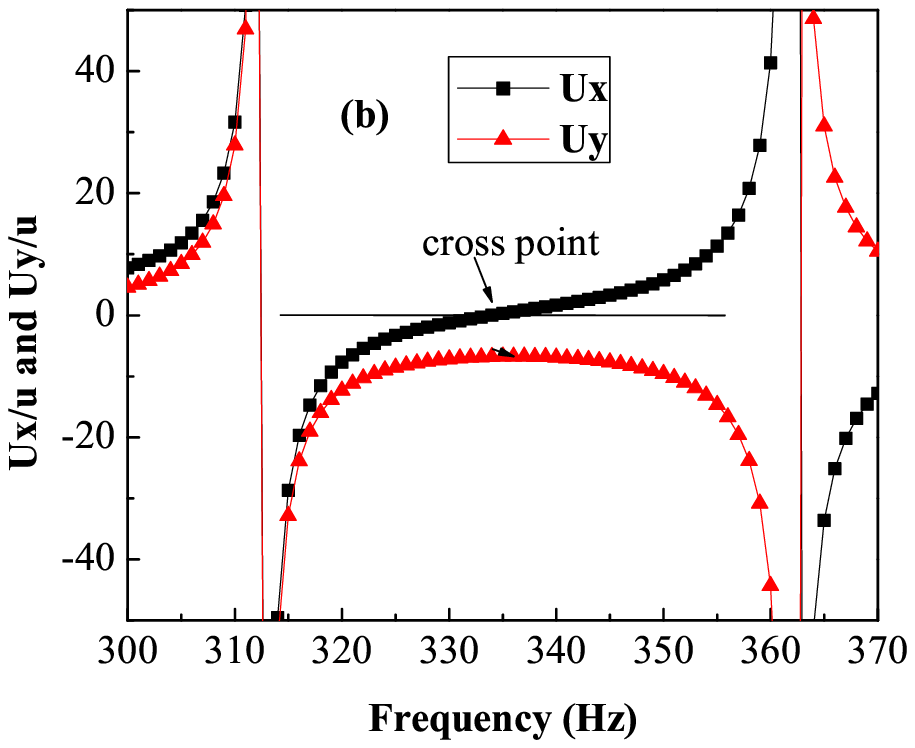}
\caption{The frequency dependence of (a) force acting on epoxy
matrix by coated resonators in $x$ and $y$ directions; (b)
displacement of cylinders in medium $2$, which is an enlargement of
the resonance part in Fig.\ref{fig:mode124}. The line with solid
squares and line with solid trigons represent $x$ and $y$ direction,
respectively. Both the force and displacement are normalized to the
displacement u in medium 1.} \label{fig:mode3force}
\end{figure}

To find out the implication of the phase reversion, we calculated
the frequency dependance of force on epoxy matrix by the coated
cylinder,
\begin{eqnarray}
F_x=\int\int_s(n_r\cos\phi\tau_{rr}+(n_\phi\cos\phi-n_r\sin\phi)\tau_{r\phi}-n_\phi\sin\phi\tau_{\phi\phi})cdld\phi\\
F_y=\int\int_s(n_r\sin\phi\tau_{rr}+(n_\phi\sin\phi+n_r\cos\phi)\tau_{r\phi}+n_\phi\cos\phi\tau_{\phi\phi})cdld\phi,
\end{eqnarray}
plotted  in Fig. \ref{fig:mode3force}.
It can be seen from Fig. \ref{fig:mode3force}(a) that between the
two resonance frequencies the force in $x$ direction is gradually
changed from positive to negative, while the force in $y$ direction
is always positive. The cross point frequency, denoted by $f_c$,
where the force in $x$ direction is zero,  is close and a little
higher than $332$Hz. On this frequency, the amplitude of the
displacement field in $y$ direction is a minimum, as is shown in
Fig. \ref{fig:mode3force}(b). Here the effective mass density (EMD)
for the coated resonator is a tensor and its $x$-component is
defined by $F^x_{23}=-\rho^x_{23}V_{23}\omega^2u$, where $V_{23}=\pi
c^2l$ is the volume of the coated cylinder. At this frequency, the
EMD for the block defined \cite{liu2005prb} as
$\rho^x_e=\phi_1\rho_1+(\phi_2+\phi_3)\rho^x_{23}$, who normalized
by $\rho_e=\phi_1\rho_1+\phi_2\rho_2+\phi_3\rho_3$, is 0.06. The
zerolike EMD for the block can give rise to an irregular wave
propagation behavior in the $x$ direction, which will be discussed
in detail below.  This irregularity can not be simulated by single
resonator in one unit cell. The nearest neighbor coupling effect of
resonators plays an important role in this resonance mode.

Both $P$-wave and $S$-wave can  exist in solids. Generally,
the energy will be trapped by excited resonators, then gradually
passed to the surrounding material in the process if the matrix's
absorption coefficient is not zero. In mode $2$ and mode $4$,
resonance appear in $x$ direction, the EMD tends to be infinity,
thus elastic wave can expected to be totally reflected by
non-dissipative material. Conversely, in mode $3$, the EMD tends to
zero in $x$ direction, so significant elastic wave transmission is expected.

\section{Finite Element Method Simulation}
To validates the results, the finite element commercial software,
COMSOL Multiphysics is used to simulate the vector displacement
field in a unit cell with a pair of long coated elliptic cylinders
immersed in epoxy, whose cross section is shown in Fig.
\ref{fig:unitpair}. Periodic boundary condition is used in the $y$
direction. The material component properties and their corresponding
dimensions are consistent with calculations in Sec. \ref{part2}.
Instead of rigid assumption for the matrix and inclusion, some
typical material parameters are assigned to the matrix and inclusion
cylinders: the Young's modulus for each components are, $E_1=4.35
GPa$, $E_2=117500 Pa$, $E_3=16 GPa$,
 the Poisson's ratios are, $\nu_1=0.368$, $\nu_2=0.469$, $\nu_3=0.44$,
and the densities are, $\rho_1=1180kg/m^3$, $\rho_2=1300kg/m^3$,
$\rho_3=11340kg/m^3$.

\subsection{Resonance Modes}
Triangular paver elements were used to  construct finite element
mesh of a unit cell with elliptic inclusions with $11173$ nodes and
$21984$ elements. The mesh size is usually set according to the
shortest wave length expected during the event
\cite{comsol2007boston}. For example, if $N_{ew}$ is the number of
elements per wavelength required for accurate modeling, $C_{min}$ is
the slowest wave speed, and $f_{max}$ the largest frequency
experienced in a frequency sweep, then the mesh  size is determined
by $\Delta_{min}=C_{min}/(N_{ew}f_{max})$, (e.g. $N_{ew}=6$ for
quadratic element shape functions and $N_{ew}=10$ for linear element
shape functions). When rubber like materials are employed, the shear
wave speed is typically the smallest and governs the mesh size
needed in the solid. In the frequency range considered in this
study, the wave-length in the matrix material is much longer than
the side length of the unit cell.

The upper and down side were set as periodic boundaries to represent
the infinite extent in $y$ direction, therefore the displacement of
epoxy matrix in this direction is zero. The left and right edges
were set
as free sides receiving and transmitting traveling waves.

\begin{figure}
\includegraphics[width=0.6\textwidth]{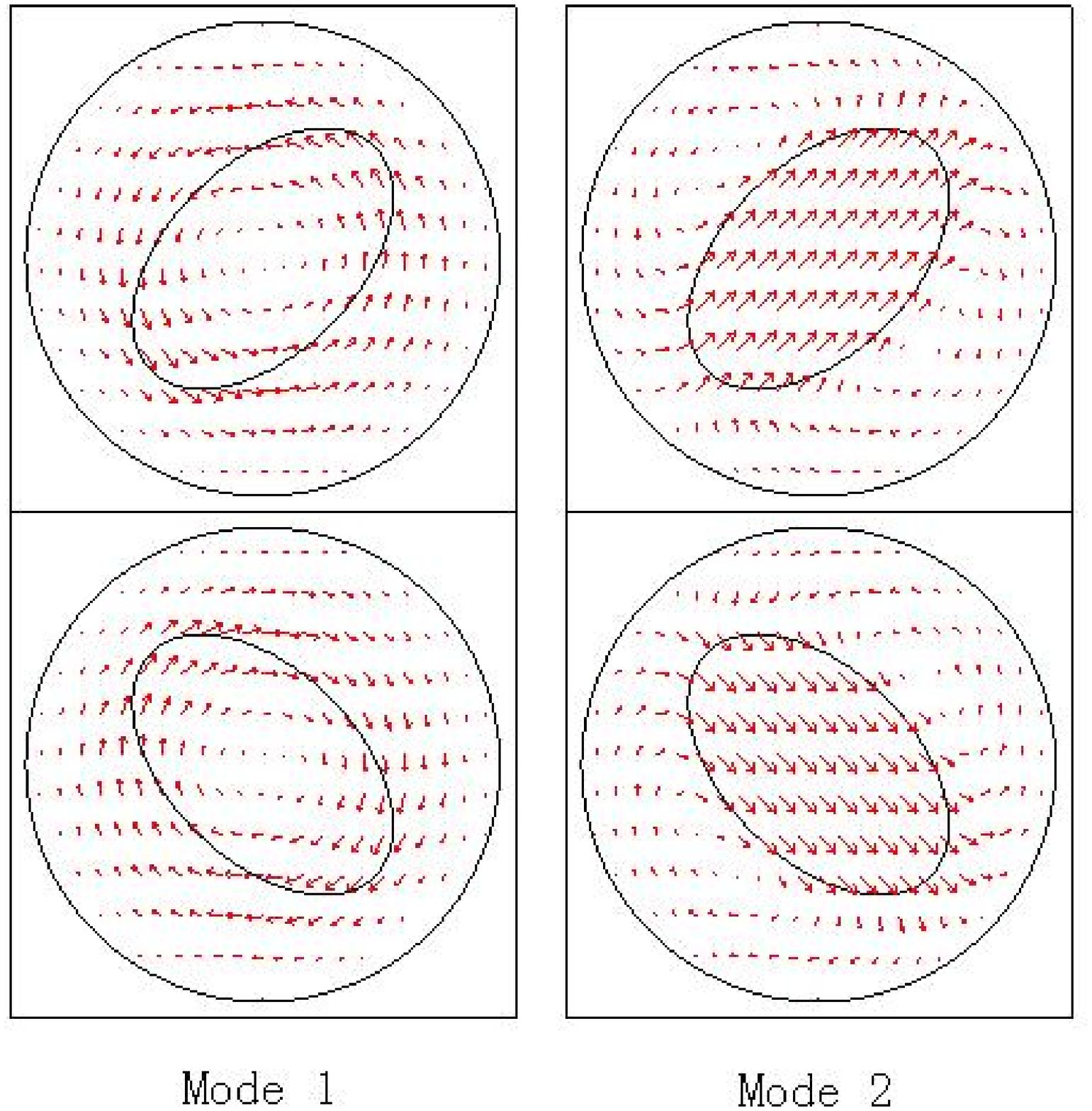}\\
\includegraphics[width=0.6\textwidth]{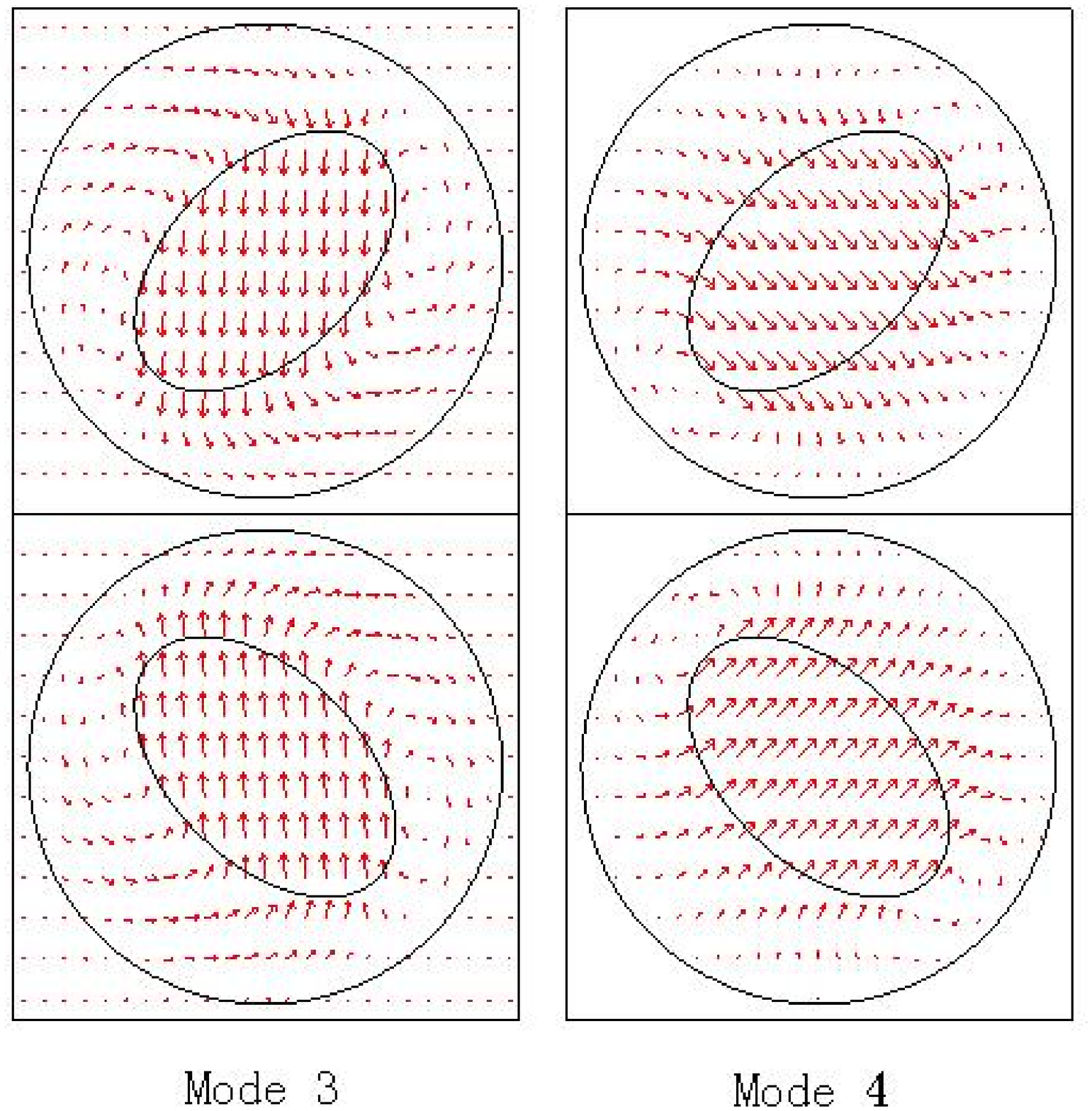}
\caption{Vibration modes in 2D ternary PC with asymmetry resonators
at four frequencies: f=258Hz, 313Hz, 332Hz and 363Hz. The direction
and length of the arrows represent the direction and amplitude of
displacement vectors, respectively.}\label{fig:allmodes}
\end{figure}

Fig. \ref{fig:allmodes} shows the resulting simulated modes. For
application reasons, higher rotational modes of the coating with an
increasing number of integer nodes are not discussed here. Each one
of the four modes related to the motion of the resonator, so all of
the resonance frequencies of these modes are expected to be in the
interesting low frequency regime. In mode $1$, the core is rotating
in the inclusion, inducing no net forces acting on the matrix, thus
make no contribution to the wave insulation. Mode $2$ and mode $4$
are formed by the substitution of the circular shaped resonators by
elliptic ones. Thus the degenerate resonance frequency in circular
case is split into two modes. The resonance frequency in mode $2$ is
$f=313Hz$, lower than the circular one ($a=b=5mm$), which is
$f=359Hz$ for the parameters used here, according to Liu $et$
$al$\cite{liu2005prb}. The    resonance frequency of mode $4$ is
$f=363Hz$,  which is higher than the circular counterpart.

Mode $3$ is  the result of the interaction between two blocks in the
unit cell, the resonance frequency is $f=332Hz$. The resonators are
placed at geometry centers of the two blocks, while the orientation
angles of the two resonators in one unit cell possess mirror
symmetry with respect to the middle plane (see Fig.
\ref{fig:unitpair}). The conjugate orientation angles for upper
block and lower blocks in unit are $45^\circ$ and $-45^\circ$,
respectively. The coupled two blocks give rise to a vibration mode,
in which two resonators move in opposite directions along layer
extension. The orientation angle can be set in the range
$\theta_0\in(0^\circ,90^\circ)$ for the upper block in Fig.
\ref{fig:unitpair}, and simultaneously set in the range
$\theta_0\in(-90^\circ,0^\circ)$ for the lower one. For the case
$\theta_0=90^\circ$ or $\theta_0=0^\circ$, this mode is identical to
that of the single block in one unit cell, merged to the mode $2$ or
mode $4$. The resonance modes predicted by analytical methods seems
to agree quite well with the finite element simulation results.

\begin{figure}
\includegraphics[scale=0.8]{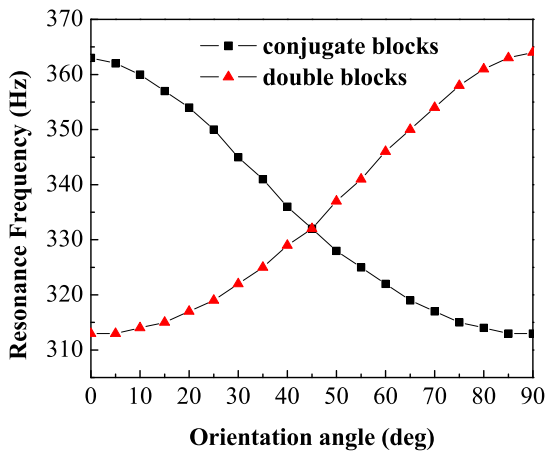}
\caption{Orientation angle dependence of resonance frequency for
conjugate blocks and double blocks in unit cell, respectively.}
\label{fig_orientang}
\end{figure}

For different conjugate orientation angles, we obtained continuously
changing resonance frequency under this coupling resonance mode
(mode 3), with the other three resonance modes remains the same.
Line with solid square in Fig. \ref{fig_orientang} shows that the
resonance frequency dependant on the orientation angle of the
elliptic cylinder. It decreases from $0^\circ$, and reaches the
minimum at $90^\circ$. At both end of the curve, mode 3 disappeared,
degenerating into that of the single block unit cell case. It also
reveals that near $45^\circ$, the slope achieves the maximum value.
Line with solid trigon in Fig. \ref{fig_orientang} shows the doubled
block in unit cell case. With the same orientation angle
$\theta_0=45^\circ$ for both upper and lower blocks, the resonance
frequency still dependent on the orientation angle, however, the
vibration mode is no longer the same. This also lead to a different
transmission property in the following section.

\subsection{Attenuation performances}
In order to provide efficient and reliable simulations for
comparison with experimental data and other applications, the
transmission properties  is needed\cite{liu2000science,
hirsekorn2006jap} and calculated as follows.

A periodic arrangement of unit cells with two coupling blocks  illustrated in Fig. \ref{fig:unitpair} is
submerged in air. An incident plane wave is traveling along the
positive $x$ direction. Upper and lower boundaries were still
exerted periodical boundary conditions. On left and right side, the
air domains were added, governed by the equation
\begin{equation}
\nabla(\frac{1}{\rho}\nabla p)+\frac{\omega^2}{\rho c^2}p=0,
\end{equation}
which is provided by Time-harmonic analysis Module in COMSOL. The
incident plane wave will encounter the air-epoxy boundary, inducing
the displacement field in solids. The displacement field in solid
domain is solved by employing Frequency response analysis Module. In
order to illustrate the similarity of the equation form in the solid
and air, we write the governing equations in the following form,
\begin{equation}
c_p^2\frac{\partial}{\partial x_j}(\frac{\partial u_i}{\partial
x_i})+c_s^2\frac{\partial}{\partial x_i}(\frac{\partial
u_j}{\partial x_i}-\frac{\partial u_i}{\partial x_j})+\omega^2
u_j=0,\qquad j=1,2,3,
\end{equation}
where $c$, $c_p$ and $c_s$ are the wave speed in air, $P$-wave speed
and $S$-wave speed in solid, respectively, and the repeated indices
are summed over $1$ to $3$. These two modules are then coupled
through domain boundaries by domain variables $p$ and $\mathbf{u}$.
The harmonic acoustic pressure in the air on the interface acts as a
boundary load $p\mathbf{n}$ to the solids. The model calculates
harmonic displacements and stresses in the solids, and then it uses
the normal acceleration $\mathbf{n}\cdot\mathbf{\ddot{u}}$ of the
solid surface for air domain boundary to ensure continuity in
acceleration.

\begin{figure}
\includegraphics[scale=0.7]{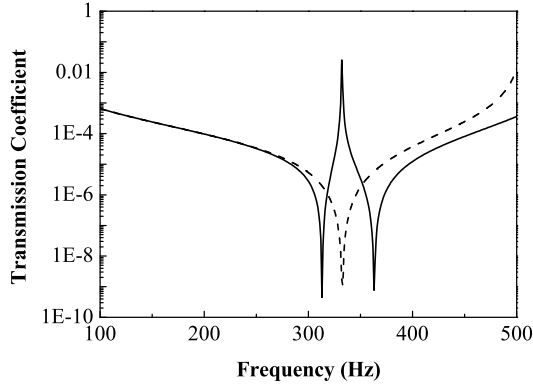}
\caption{Wave transmission spectrum in the low frequency region
obtained from FEM simulations of a single layer composed by unit
cells, the frequency resolution is $1Hz$. Solid and dashed lines
represent unit cell with conjugate blocks and double blocks,
respectively.} \label{fig:orientang}\label{spectrum}
\end{figure}

Fig. \ref{spectrum} shows the elastic wave transmission spectrum
from $100Hz$ to $500Hz$. Single frequency harmonic waves were
employed in the simulation, and the minimum step is $1Hz$. In the
calculation, the absorption coefficient of the materials was taken
to be zero, and transmission coefficient calculated for intensity of
waves. The two dips and one peak observed in this figure are
associated with the motion of inner resonator. The results from
modal analysis are in good agreement with this transmission
spectrum, in terms of the frequency positions of the two dips and
the peak. The two dips, appearing at $313Hz$ and $363Hz$, are
corresponding to resonance vibration in Mode $2$ and Mode $4$. Mode
$1$  can not be observed in this figure, because the resonator
remains symmetric in some extent. In the special case of elliptical
resonators, whose center of  mass away from its geometric center or
the main axes are off the direction of wave propagation, rotational
resonance can be expected to activate, bringing with an attenuation
dip. For Mode $3$, with conjugate orientation angles $45^\circ$ and
$-45^\circ$, a peak at $332Hz$ is observed. This peak appears as an
manifestation of the zero effective mass density for the block in
$x$ direction. The peak value is in fact $1$, the lower value in the
figure is an artifact due to the larger bin size in frequency used
in the calculation. Accurate results around the resonance frequency
will be given below.

No phase shift can be observed in the transmission
spectrum. The two dips in transmission spectrum have been discussed by
C. Goffaux \cite{goffaux2002prl} and M. Hirsekorn
\cite{hirsekorn2006jap}. The resonance transmission peak and its
corresponding mode is new.

\begin{figure}
\includegraphics[scale=0.7]{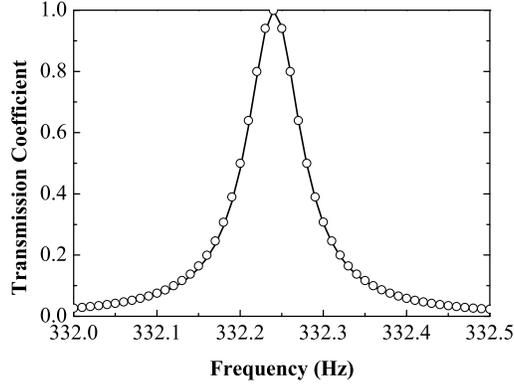}
\caption{Enlargement of the peak of the transmission spectrum in
Fig. \ref{spectrum}, the solid line represents the transmission peak
in frequency step $0.01Hz$, the circles are the fits to the
calculated data with  Eq. (\ref{trans}).} \label{zspectrum}
\end{figure}

At the resonance frequency the peak value is close to $100\%$. Away
from the resonance frequency the transmittance peak can be fitted
neatly by the Lorentz form
\begin{equation}
T(\omega)=t_0\frac{\omega^2\Gamma^2}{(\omega^2-\omega_0^2)^2+\omega^2\Gamma^2}
\label{trans}
\end{equation}
where $\omega_0$ is the resonance frequency, $\Gamma$ is the full
width at half maximum of the resonance, and $t_0$ the normalization
constant. In Fig. \ref{zspectrum} we plot the transmittance around the resonance frequency and the
fit to a Lorentz curve, in this plot the frequency step is taken to be $0.01Hz$.
This result shows that the slab is effectively absent to the
incident wave at a frequency near $332$Hz, as predicted at the end
of Sec. \ref{part2}. We denote this frequency by $f_z$. It is worth
noting that the $f_z$ is a little higher than $f_c$, which in turn
is slightly higher than the resonance frequency $332$Hz. And all of
the three are located in open interval $(332Hz, 333Hz)$. When
gradually tend to $f_z$ from the resonance frequency, $\rho^x_e$
turns to zero, in other words, the effective thickness of the PC
layer becomes smaller until zero, so that the attenuation effect is
weakened.

In comparison, we also calculated the transmittance of a system with
a unit cell where the two elliptic cylinders in the same orientation
($\theta_0=45^\circ$), which is a case of double blocks in unit cell
of a simple PC. The result is shown in Fig. \ref{spectrum} as dashed
line. It can be seen that
 the resonance modes $2$ and $4$  disappear completely, and mode $3$ at frequency $332$Hz (the cross point in Fig. \ref{fig_orientang}) turns to be a transmission dip.
 Resonance mode analysis indicates  that the whole rigid matrix takes part in to partially balance the momentum in
$y$ direction induced by the cylinders, while cylinder resonance
mainly exits  in $x$ direction. This lead to large EMD in $x$
direction, and hence significant reflection for non-dissipative
material.
 Actually, the minimum repeating unit of the system is the single cell. As investigated by Hirsekorn $et$ $al$\cite{hirsekorn2006jap},
 it makes no difference when the unit cell is doubled.

\section{Conclusion}
In summary, the influence of the local asymmetry on the properties
of the PCs at the low frequency in 2D has been studied. The
analytical method used here are able to identify rotational mode,
and the new mode whose vibration is perpendicular to the incident
direction. We compared the results with the ones from finite element
method, showing good agreement with each other. Transmission
characteristic of a single layer PC are also calculated. The two
dips are elastic wave attenuation corresponding to the common
resonance modes, while the peak corresponds to the new resonance
mode. Moreover, the peak can be tuned by simultaneously rotating the
orientation angle of elliptical resonators of coupling blocks in a
unit cell. This property provides a simple way to tune the resonance
transmittance in locally resonant composite materials, leading to
further opportunities for practical applications of the PCs.


\end{document}